\documentclass[12pt]{article}

\begin{document}

\title{Moyal-Weyl Star-products as Quasiconformal Mappings}
\author{Tadafumi Ohsaku}

\date{\today}

\maketitle



In the methods of deformation quantization~[1-7], 
the Moyal-Weyl star product $f\star g$ is quite often used in mathematical/theoretical physics.
Let us consider the case that the base ring of a theory is a two-dimensional 
real number field ${\bf R}^{2}$.
In the framework of two-dimensional quantum field theory~[8], 
the coordinate system $(x,y)\in{\bf R}^{2}$
will be converted into a complex coordinates $(z,\bar{z})\in{\bf C}$
with the definition $z\equiv x+iy$, 
and the methods of complex analysis is introduced in various calculations of the theory.
We will examine the star product $f(z,\bar{z})\star g(z,\bar{z})$.
The main argument of this paper is that,
the deformation-quantization procedure can introduce the quantum effect 
if and only if not both of $f$ and $g$ are holomorphic, 
namely including the deformation of complex structure
of the domains of functions $f$ and $g$
( or, a pair of holomorphic and antiholomorphic functions ),  
and ( sometimes ) the star product becomes a quasiconformal mapping.
The ultimate purpose of this paper is to make the road toward a Teichm\"{u}ller theory
of the Moyal-Weyl star product through the quasiconformal mappings as liftings of
universal covering surfaces of any Riemann surfaces.

\vspace{10mm}

First, we give the definition of the Moyal-Weyl star product
as follows: 
\begin{eqnarray}
&& w = {\cal F}(z,\bar{z}), \nonumber \\
&& {\cal F}(z,\bar{z}) \equiv f_{1}(z,\bar{z})\star f_{2}(z,\bar{z}), \quad 
( z\in D, \, \bar{z}\in D^{*})  \nonumber \\
&& \star \equiv \exp\Bigg[i\hbar\Bigl( 
\overleftarrow{\partial}_{z}\overrightarrow{\partial}_{\bar{z}}
-\overleftarrow{\partial}_{\bar{z}}\overrightarrow{\partial}_{z}  
\Bigr)\Bigg],  \nonumber \\
&& \overline{{\cal F}(z,\bar{z})} = \overline{f_{1}(z,\bar{z})}\star\overline{f_{2}(z,\bar{z})}, \quad
( {\rm when} \, \bar{\hbar} = \hbar ),   \nonumber \\
&& {\cal G}(z,\bar{z}) \equiv f_{2}(z,\bar{z})\star f_{1}(z,\bar{z}),   \nonumber \\
&& \overline{{\cal G}(z,\bar{z})} = \overline{f_{2}(z,\bar{z})}\star\overline{f_{1}(z,\bar{z})}.
\end{eqnarray}
Here, we choose the operator of the star product $\star$ to generate
the Poisson brackets in the canonical form.
The algebra of the products are possibly noncommutative, i.e., 
${\cal F}\ne{\cal G}$, $\overline{\cal F}\ne\overline{\cal G}$.
We consider only the ( planer ) domain of $z$ as $\widehat{\bf C}$ ( the Riemann sphere ), 
${\bf C}$ ( the complex $z$-plane ) and ${\bf H}$ ( the upper half of the $z$-plane )
due to the uniformization theorem.

\vspace{10mm}

Our main statement of this paper is summarized in the following proposition:
{\bf Proposition.1}
{\it The Moyal-Weyl star product ${\cal F}=f\star g$ becomes a quasiconformal mapping 
under suitable choices for $f$ and $g$. 
To make sense the procedure of deformation quantization, 
$f$ and $g$ must not be ( conformal ) holomorphic functions of $z$.}
In other words, $f$ and $g$ have to have the nature of 
discrepancies from conformal ( biholomorphic ) character
( or, a pair of holomorphic and antiholomorphic functions at least )
for obtaining the deformation quantization of the Moyal-Weyl star product.
Hence in that case, the Moyal-Weyl deformation quantization procedure 
has deep connection with the deformation of complex structure,
and then the situation might relate the star products to 
the Moduli and Teichm\"{u}ller spaces of Riemann surfaces.
In this paper, we will observe them by several examples.

\vspace{10mm}

The definition of the quasiconformal mapping is given as follows~[9]:
{\bf Definition.1} $f$, a homeomorphism of a domain $D$, is called as a quasiconformal mapping, when
(i) $f$ is partially differentiable almost everywhere of $D$, 
(ii) $|f_{\bar{z}}|\le k|f_{z}|$ ( $0\le k < 1$ ) is satisfied over $D$.

\vspace{10mm}

A quasiconformal map $f$ on a domain $D$ satisfies the following conditions~[9]:

(I) The partial derivatives $f_{z}$ and $f_{\bar{z}}$ are square integrable:
\begin{eqnarray}
\int_{D}|f_{z}|^{2} < +\infty, \quad
\int_{D}|f_{\bar{z}}|^{2} < +\infty. 
\end{eqnarray}

(II) $f_{z}$ and $f_{\bar{z}}$ are partial derivatives in the sense of distribution:
\begin{eqnarray}
\int_{D}dxdy f_{z}\varphi &=& -\int_{D}dxdy f\varphi_{z},  \nonumber \\
\int_{D}dxdy f_{\bar{z}}\varphi &=& -\int_{D}dxdy f\varphi_{\bar{z}}, 
\end{eqnarray}
where, $\varphi$ is a smooth function with a compact support over $D$.

(III) $f_{z}\ne 0$ is satisfied over $D$.

\vspace{10mm}

We consider the case where $f_{1}(z,\bar{z})$, $f_{2}(z,\bar{z})$ and 
${\cal F}(z,\bar{z})=f_{1}(z,\bar{z})\star f_{2}(z,\bar{z})$,
and we introduce the following Beltrami equations:
\begin{eqnarray}
& & \partial_{\bar{z}}f_{1} = \mu_{f_{1}}\partial_{z}f_{1}, \quad
\partial_{\bar{z}}f_{2} = \mu_{f_{2}}\partial_{z}f_{2},  \quad
\partial_{\bar{z}}{\cal F} = \mu_{\cal F}\partial_{z}{\cal F},  \nonumber \\ 
& & ( \mu_{f_{1}}, \mu_{f_{2}}, \mu_{\cal F} \in {\bf C} ). 
\end{eqnarray}
Then, if $f_{1}$ and $f_{2}$ satisfy these equations, the Poisson bracket will be expressed as follows:
\begin{eqnarray}
\{f_{1},f_{2}\}^{P.B.}_{z,\bar{z}} &\equiv& 
\frac{\partial f_{1}}{\partial z}\frac{\partial f_{2}}{\partial\bar{z}} -
\frac{\partial f_{1}}{\partial\bar{z}}\frac{\partial f_{2}}{\partial z}  
= \bigl(\mu_{f_{2}}-\mu_{f_{1}}\bigr)\frac{\partial f_{1}}{\partial z}\frac{\partial f_{2}}{\partial z}.
\end{eqnarray}
Hence, the Poisson bracket vanishes at the region $\mu_{f_{1}}=\mu_{f_{2}}$.
Usually, the star product is interpreted as an expansion of the power series of $\hbar$
of ${\cal F}=f_{1}\star f_{2}$.
Our star product ${\cal F}$ may be expanded by 
$\hbar$ or the Beltrami coefficients $\mu_{f_{1}}$ and $\mu_{f_{2}}$:
\begin{eqnarray}
{\cal F} &=& {\cal F}^{(0)} + \hbar{\cal F}^{(1)} + \hbar^{2}{\cal F}^{(2)} + \cdots   \nonumber \\
&=& \tilde{\cal F}^{(0)} + \mu_{f_{1}}\tilde{\cal F}^{(1)}_{1} + \mu_{f_{2}}\tilde{\cal F}^{(1)}_{2} 
+ \mu^{2}_{f_{1}}\tilde{\cal F}^{(2)}_{1} + \mu_{f_{1}}\mu_{f_{2}}\tilde{\cal F}^{(2)}_{12}
+ \mu^{2}_{f_{2}}\tilde{\cal F}^{(2)}_{2}
+ \cdots,    \nonumber \\
& & 
\end{eqnarray}
and the Beltrami coefficient $\mu_{\cal F}$ of the equation
$\partial_{\bar{z}}{\cal F}=\mu_{\cal F}\partial_{z}{\cal F}$ is also expanded as
\begin{eqnarray}
\mu_{\cal F} &=& \mu^{(0)}_{\cal F} + \hbar\mu^{(1)}_{\cal F} + \hbar^{2}\mu^{(2)}_{\cal F} + \cdots.
\end{eqnarray}
Hence, the star product ${\cal F}$ is given as a superposition of ${\cal F}^{(n)}$,
or that of $\tilde{\cal F}^{(n)}$ ( $n\in{\bf Z}$, $0\le n < +\infty$ ),
and this indicates a deep relation between the Beltrami coefficients and $\hbar$ in the star product.

\vspace{10mm}

The affine mappings $f_{1}$ and $f_{2}$ defined as follows 
will become examples of quasiconformal mappings:
\begin{eqnarray}
& & f_{1}(z,\bar{z}) = a_{1}z + b_{1}\bar{z} + c_{1}, \quad 
f_{2}(z,\bar{z}) = a_{2}z + b_{2}\bar{z} + c_{2},   \nonumber \\
& & ( a_{j},b_{j},c_{j}\in{\bf C}, \, |a_{j}| > |b_{j}|, \, j=1,2 ).
\end{eqnarray}
From the Beltrami equations of them, one finds
\begin{eqnarray}
\partial_{\bar{z}}f_{1} = \mu_{f_{1}}\partial_{z}f_{1} &\longrightarrow& \mu_{f_{1}} = \frac{b_{1}}{a_{1}},  \nonumber \\
\partial_{\bar{z}}f_{2} = \mu_{f_{2}}\partial_{z}f_{2} &\longrightarrow& \mu_{f_{2}} = \frac{b_{2}}{a_{2}}.
\end{eqnarray} 
Hence
\begin{eqnarray}
{\cal F} &=& f_{1}\star f_{2} 
= f_{1}f_{2} + i\hbar(a_{1}b_{2}-b_{1}a_{2}) 
= f_{1}f_{2} - i\hbar(\mu_{f_{1}}-\mu_{f_{2}})a_{1}a_{2}.
\end{eqnarray}
${\cal F}$ is holomorphic as a function of $\mu_{f_{1}}$, $\mu_{f_{2}}$,
$a_{1}$, $c_{1}$, $a_{2}$ and $c_{2}$, while not holomorphic as a function of $z$ and $\bar{z}$. 
The function ${\cal F}$ itself has the form where it is holomorphic over
${\bf C}^{2}\equiv{\bf C}_{\mu_{f_{1}}}\otimes{\bf C}_{\mu_{f_{2}}}$ 
( $\mu_{f_{j}}\in{\bf C}_{\mu_{f_{j}}}$, $j=1,2$ ),
or over the two dimensional Osgood space~[10] 
$\widehat{\bf C}^{2}\equiv\widehat{\bf C}_{\mu_{f_{1}}}\otimes\widehat{\bf C}_{\mu_{f_{2}}}$
( $\widehat{\bf C}_{\mu_{f_{j}}} \equiv {\bf C}_{\mu_{f_{j}}}+\{\infty\}$, $j=1,2$ ).
${\cal F}$ is convergent 
in the whole of ${\bf C}^{2}$ as a function of $\mu_{f_{j}}$ ( $j=1,2$ ).
We find that, when $\mu_{f_{1}}=\mu_{f_{2}}$ ( the case $f_{1}=f_{2}$ is a specific example of it ), 
the effect of deformation quantization cannot be introduced.
The situation where $f_{1}$ and $f_{2}$ are conformal ( biholomorphic ) on $z$, 
namely $\mu_{f_{1}}=\mu_{f_{2}}=0$, the deformation quantization cannot be done.
The criterion for quasiconformal mappings restricts
the Beltrami coefficients $\mu_{f_{1}}$ ( $j=1,2$ ) to satisfy
\begin{eqnarray}
0\le |\mu_{f_{1}}| < 1, \quad 0\le |\mu_{f_{2}}| < 1.
\end{eqnarray}
Thus, we consider a poly-unit-disc ${\cal D}_{\mu_{f_{1}}}\otimes{\cal D}_{\mu_{f_{2}}}$ 
with centre $(\mu_{f_{1}},\mu_{f_{2}})=(0,0)$ 
for the domain of the space of $(\mu_{f_{1}},\mu_{f_{2}})$. This open domain is noncompact. 
Furthermore, by a suitable combination of a rotation $\mu_{f_{j}}\to e^{i\theta}\mu_{f_{j}}$
( $\theta\in{\bf R}$, $\theta\ne 0$ ) and a dilatation $\mu_{f_{j}}\to \lambda\mu_{f_{j}}$
( $0<\lambda<1/|\mu_{f_{j}}|$, $\lambda\ne 1$ ), we can always make $\mu_{f_{1}}=\mu_{f_{2}}$
and in that case the effect of the deformation quantization vanishes.
These transformations correspond to the variations of parameters $a_{j}$ and $b_{j}$ ( $j=1,2$ )
of the affine maps. We show the invariance of $|\mu_{f_{j}}|$ under a conformal 
( biholomorphic ) map $\varphi: z\to\varphi(z)$:
\begin{eqnarray}
& & \varphi: a_{j}z+b_{j}\bar{z}+c_{j} \longrightarrow a_{j}\varphi(z)+b_{j}\overline{\varphi(z)}+c_{j},   \nonumber \\
& & \partial_{z}f_{j}(\varphi(z),\overline{\varphi(z)}) = a_{j}\frac{\partial\varphi}{\partial z},   \nonumber \\ 
& & \partial_{\bar{z}}f_{j}(\varphi(z),\overline{\varphi(z)}) = b_{j}\frac{\partial\overline{\varphi}}{\partial\bar{z}},   \nonumber \\
& & |\mu_{f_{j}}| = 
\Bigg|\frac{\partial_{\bar{z}}f_{j}(\varphi,\overline{\varphi})}
{\partial_{z}f_{j}(\varphi,\overline{\varphi})}\Bigg| 
= \Bigg|\frac{b_{j}\partial_{\bar{z}}\overline{\varphi}}{a_{j}\partial_{z}\varphi}\Bigg|
= \Bigg|\frac{b_{j}}{a_{j}}\Bigg|.
\end{eqnarray}

\vspace{10mm}

Next, we show another example where $f_{1}$, $f_{2}$ and ${\cal F}$ can become quasiconformal:
\begin{eqnarray}
f_{1}(z,\bar{z}) &=& e^{i\alpha_{1}z}e^{i\beta_{1}\bar{z}} 
= e^{i(\alpha_{1}+\beta_{1})x-(\alpha_{1}-\beta_{1})y}, \nonumber \\
f_{2}(z,\bar{z}) &=& e^{i\alpha_{2}z}e^{i\beta_{2}\bar{z}}
= e^{i(\alpha_{2}+\beta_{2})x-(\alpha_{2}-\beta_{2})y},  \nonumber \\ 
& & \qquad \qquad 
( \alpha_{1}, \alpha_{2}, \beta_{1}, \beta_{2} \in {\bf C}, \alpha_{1}, \alpha_{2} \ne 0) \nonumber  \\
{\cal F}(z,\bar{z}) &=& f_{1}(z,\bar{z})\star f_{2}(z,\bar{z})
= {\cal F}^{(0)} + \hbar{\cal F}^{(1)} + {\cal O}(\hbar^{2}),  \nonumber \\
{\cal F}^{(0)} &=& f_{1}f_{2} = e^{i(\alpha_{1}+\alpha_{2})z}e^{i(\beta_{1}+\beta_{2})\bar{z}}, \nonumber  \\
{\cal F}^{(1)} &=& i\{f_{1},f_{2}\}^{P.B.}_{z,\bar{z}} 
= -i(\alpha_{1}\beta_{2}-\beta_{1}\alpha_{2})f_{1}f_{2} = i(\mu_{f_{1}}-\mu_{f_{2}})f_{1}f_{2},  \nonumber  \\
{\cal F} &=& e^{-i\hbar(\alpha_{1}\beta_{2}-\beta_{1}\alpha_{2})}f_{1}f_{2}.
\end{eqnarray}
The functions $f_{j}=e^{i(\alpha z+\beta\bar{z})}$ can be regarded ( more precisely, include ) 
as a result of taking affine mapping $z\to(K+1)z/2+(K-1)\bar{z}/2$ 
in a conformal ( holomorphic ) function $e^{iz}$.
In other words, $f_{j}$ ( $j=1,2$ ) become conformal when $\beta_{j}=0$ ( $j=1,2$ ).
Clearly, $f_{j}$ are square integrable and partially differentiable almost everywhere of a domain $D$.
Because $\alpha_{j}\ne0$, $\partial_{z}f_{j}\ne0$ ( $j=1,2$ ) are satisfied over any $D$ of ${\bf C}$.
Hence the Beltrami equations for $f_{j}$ can be defined.
$f_{j}$ ( $j=1,2$ ) induce the maps $f_{1},f_{2};\widehat{\bf C}\to\widehat{\bf C}$,
and $f_{1},f_{2};{\bf C}\to{\bf C}-\{0\}$.
We have observed that, the effect of the deformation quantization is introduced 
as the phase factor to the product $f_{1}f_{2}$ of this example,
and ${\cal F}$ takes the similar functional structure with $f_{j}$.
${\cal F}$ is absolutely convergent at $0\le |\hbar|<\infty$.
It is an interesting fact that, 
the effect of deformation quantization disappears when $\mu_{f_{1}}=\mu_{f_{2}}$. 
In this case the Poisson bracket also identically vanishes.
If we use the Beltrami equations, we find the following relations between 
$\mu_{f_{j}}$, $\alpha_{j}$ and $\beta_{j}$ ( $j=1,2$ ): 
\begin{eqnarray}
\partial_{\bar{z}}f_{1} = \mu_{f_{1}}\partial_{z}f_{1}  &\longrightarrow& \mu_{f_{1}} = \frac{\beta_{1}}{\alpha_{1}}, \nonumber  \\
\partial_{\bar{z}}f_{2} = \mu_{f_{2}}\partial_{z}f_{2}  &\longrightarrow& \mu_{f_{2}} = \frac{\beta_{2}}{\alpha_{2}}.  
\end{eqnarray}
Thus, one obtains the following results from $\partial_{\bar{z}}{\cal F}=\mu_{{\cal F}}\partial_{z}{\cal F}$:
\begin{eqnarray}
\mu_{\cal F} &=& \frac{\beta_{1}+\beta_{2}}{\alpha_{1}+\alpha_{2}} 
= \frac{\mu_{f_{1}}\alpha_{1}+\mu_{f_{2}}\alpha_{2}}{\alpha_{1}+\alpha_{2}},  \nonumber \\
{\cal F} &=& e^{i(\alpha_{1}+\alpha_{2})z}e^{i(\mu_{f_{1}}\alpha_{1}+\mu_{f_{2}}\alpha_{2})\bar{z}}
e^{i\hbar(\mu_{f_{1}}-\mu_{f_{2}})\alpha_{1}\alpha_{2}}.
\end{eqnarray}
$z=0$ cannot be a fixed point of $f_{1}$, $f_{2}$ and ${\cal F}$
( Usually, $0$, $1$ and $\infty$ will be chosen as fixed points to solve the Beltrami equation ).
The star product ${\cal F}$ is analytic ( holomorphic ) on both in $\mu_{f_{1}}$ and $\mu_{f_{2}}$,
and satisfies the Cauchy-Riemann equations:
\begin{eqnarray}
\partial_{\overline{\mu_{f_{1}}}}{\cal F} = 0, \quad 
\partial_{\overline{\mu_{f_{2}}}}{\cal F} = 0.
\end{eqnarray}
Moreover,
\begin{eqnarray}
& & \partial_{\mu_{f_{1}}}\partial_{\overline{\mu_{f_{1}}}}{\cal F} = 0,  \quad 
\partial_{\mu_{f_{1}}}\partial_{\overline{\mu_{f_{2}}}}{\cal F} = 0,  \quad
\partial_{\mu_{f_{2}}}\partial_{\overline{\mu_{f_{1}}}}{\cal F} = 0,  \quad 
\partial_{\mu_{f_{2}}}\partial_{\overline{\mu_{f_{2}}}}{\cal F} = 0.  \nonumber \\
& & 
\end{eqnarray}
Therefore, if we consider the star product ${\cal F}$ as a function of $\mu_{f_{1}}$ and $\mu_{f_{2}}$,
it will be expressed in the following form through the Cauchy theorem of 
the case of several complex variables~[10,11],
\begin{eqnarray}
{\cal F}(\mu_{f_{1}},\mu_{f_{2}}) = \int_{C_{1}}\frac{d\zeta_{1}}{2\pi i}\int_{C_{2}}\frac{d\zeta_{2}}{2\pi i}
\frac{{\cal F}(\zeta_{1},\zeta_{2})}{(\zeta_{1}-\mu_{f_{1}})(\zeta_{2}-\mu_{f_{2}})}.
\end{eqnarray}
Here, $C_{1}$ and $C_{2}$ are appropriate integration paths inside ${\bf C}$.
$\zeta_{1}$ and $\zeta_{2}$ can take the values over ${\bf C}$,
because they do not have the condition of quasiconformal maps.  
Under a conformal map $\phi:z\to\phi(z)$, an absolute value of Beltrami coefficient $|\mu_{F}|$
of a function $F(z,\bar{z})=e^{i\alpha z}e^{i\beta\bar{z}}$ is invariant:
\begin{eqnarray}
\partial_{z}F(\phi(z),\overline{\phi(z)}) &=& 
i\alpha(\partial_{z}\phi(z))F(\phi(z),\overline{\phi(z)}), \nonumber \\
\partial_{\bar{z}}F(\phi(z),\overline{\phi(z)}) 
&=& i\beta(\partial_{\bar{z}}\overline{\phi(z)})F(\phi(z),\overline{\phi(z)}), \nonumber \\
|\mu_{F}| 
&=& \Bigg|\frac{i\beta(\partial_{\bar{z}}\overline{\phi(z)})F(\phi(z),\overline{\phi(z)})}
{i\alpha(\partial_{z}\phi(z))F(\phi(z),\overline{\phi(z)})}\Bigg| = \Bigg|\frac{\beta}{\alpha}\Bigg|.
\end{eqnarray}
Hence, we confirm that, from the form of the functions $f_{1}$, $f_{2}$ and ${\cal F}$, 
$|\mu_{f_{1}}|$, $|\mu_{f_{2}}|$ and $|\mu_{\cal F}|$ are invariant under
a conformal map $\phi:z\to\phi(z)$.
The invariance of $|\mu_{f_{j}}|$ under $f_{j}\circ\phi$ over a $D$
guarantees us to utilize the uniformization theorem.

\vspace{10mm}

The associativity of the star product $(f_{1}\star f_{2})\star f_{3}= f_{1}\star(f_{2}\star f_{3})$
is satisfied. For example,
\begin{eqnarray}
f_{j}(z,\bar{z}) &=& e^{i\alpha_{j}z}e^{i\beta_{j}\bar{z}}, \, ( j = 1,2,3 ), \nonumber \\ 
{\cal F}(z,\bar{z}) &=& f_{1}(z,\bar{z})\star f_{2}(z,\bar{z})\star f_{3}(z,\bar{z})   \nonumber \\
&=& (f_{1}\star f_{2})\star f_{3}= f_{1}\star(f_{2}\star f_{3})   \nonumber \\
&=& e^{i(\alpha_{1}+\alpha_{2}+\alpha_{3})z}e^{i(\beta_{1}+\beta_{2}+\beta_{3})\bar{z}} \nonumber \\
& & \times e^{-i\hbar[
(\alpha_{1}\beta_{2}-\beta_{1}\alpha_{2})+
(\alpha_{2}\beta_{3}-\beta_{2}\alpha_{3})+
(\alpha_{1}\beta_{3}-\beta_{1}\alpha_{3})]},
\end{eqnarray}
while the operation of the star products do not commute: 
$f_{1}\star f_{2}\star f_{3}\ne f_{2}\star f_{3}\star f_{1}$, etc.
By using $\partial_{\bar{z}}f_{j}=\mu_{f_{j}}\partial_{z}f_{j}$ ( $j=1,2,3$ ), one finds
\begin{eqnarray}
{\cal F} &=& e^{i(\alpha_{1}+\alpha_{2}+\alpha_{3})z}e^{i(\beta_{1}+\beta_{2}+\beta_{3})\bar{z}}  \nonumber \\
& & \times e^{i\hbar[
(\mu_{f_{1}}-\mu_{f_{2}})\alpha_{1}\alpha_{2}+
(\mu_{f_{2}}-\mu_{f_{3}})\alpha_{2}\alpha_{3}+
(\mu_{f_{1}}-\mu_{f_{3}})\alpha_{1}\alpha_{3} ]}   \nonumber \\
&=& e^{i(\alpha_{1}+\alpha_{2}+\alpha_{3})z} \nonumber \\
& & \times 
e^{i[\alpha_{1}\bar{z}+\hbar(\alpha_{1}\alpha_{2}+\alpha_{1}\alpha_{3})]\mu_{f_{1}}}
e^{i[\alpha_{2}\bar{z}-\hbar(\alpha_{1}\alpha_{2}-\alpha_{2}\alpha_{3})]\mu_{f_{2}}}
e^{i[\alpha_{3}\bar{z}-\hbar(\alpha_{2}\alpha_{3}+\alpha_{1}\alpha_{3})]\mu_{f_{3}}}.  \nonumber \\
&&
\end{eqnarray}
In this form, ${\cal F}$ is holomorphic, has no zero point, $C^{\infty}$-class function,
and absolutely convergent over 
${\bf C}^{3}={\bf C}_{\mu_{f_{1}}}\otimes{\bf C}_{\mu_{f_{2}}}\otimes{\bf C}_{\mu_{f_{3}}}$ as a function of 
$\mu_{f_{1}}$, $\mu_{f_{2}}$ and $\mu_{f_{3}}$.
Similar to the example of the affine mapping discussed above, 
if we want to make both $f_{j}$ and the star product ${\cal F}$ as quasiconformal mappings, 
the condition of it gives the restrictions $0\le |\mu_{f_{j}}| < 1$ ( $j=1,2,3$ )
with $0\le |\mu_{\cal F}|=|\sum^{3}_{j=1}\mu_{f_{j}}\alpha_{j}/\sum^{3}_{j=1}\alpha_{j}|<1$, 
and this gives a poly-unit-disc as the ( open ) domain of
${\cal F}$ as a function of $\mu_{f_{j}}$ ( $j=1,2,3$ ). 
The region of $\mu_{f_{1}}=\mu_{f_{2}}=\mu_{f_{3}}$ in this poly-unit-disc, 
the quantization procedure of the Moyal-Weyl star product will vanish in ${\cal F}$.
For example, ${\cal F}$ becomes $(f_{1}f_{2})\star f_{3}$ when $\mu_{f_{1}}=\mu_{f_{2}}\ne\mu_{f_{3}}$.

\vspace{10mm}

From these results, we conclude that, in general, 
\begin{eqnarray}
&& {\cal F} = f_{1}\star f_{2}\star \cdots \star f_{n}, \quad 
f_{j} = e^{i\alpha_{j}z}e^{i\beta_{j}\bar{z}}, \, ( j = 1,2,\cdots,n ), \nonumber \\ 
&& \partial_{\bar{z}}f_{j} = \mu_{f_{j}}\partial_{z}f_{j}, \quad
\mu_{f_{j}} = \frac{\beta_{j}}{\alpha_{j}}, \quad  
\mu_{\cal F} = \frac{\sum^{n}_{j=1}\beta_{j}}{\sum^{n}_{j=1}\alpha_{j}},
\end{eqnarray}
and
\begin{eqnarray}
\frac{\partial{\cal F}}{\partial\overline{\mu_{f_{j}}}} &=& 0, \quad ( j = 1,\cdots,n ),  \nonumber \\
{\cal F}(\mu_{f_{1}},\cdots,\mu_{f_{n}}) &=& \int_{C_{1}}\frac{d\zeta_{1}}{2\pi i}\cdots \int_{C_{n}}\frac{d\zeta_{n}}{2\pi i}
\frac{{\cal F}(\zeta_{1},\cdots,\zeta_{n})}{(\zeta_{1}-\mu_{f_{1}})\cdots(\zeta_{n}-\mu_{f_{n}})},   \nonumber \\
\frac{\partial^{m_{1}+\cdots+m_{n}}{\cal F}}{\partial^{m_{1}}_{\mu_{f_{1}}}\cdots\partial^{m_{n}}_{\mu_{f_{n}}}} 
&=& (m_{1}!\cdots m_{n}!) \nonumber \\
& & \times \int_{C_{1}}\frac{d\zeta_{1}}{2\pi i}\cdots \int_{C_{n}}\frac{d\zeta_{n}}{2\pi i}
\frac{{\cal F}(\zeta_{1},\cdots,\zeta_{n})}{(\zeta_{1}-\mu_{f_{1}})^{m_{1}+1}\cdots(\zeta_{n}-\mu_{f_{n}})^{m_{n}+1}}.    \nonumber \\
&&
\end{eqnarray}
Thus, for $w={\cal F}(\mu_{f_{1}},\cdots,\mu_{f_{n}})$,
we consider $(\mu_{f_{1}},\cdots,\mu_{f_{n}})\in{\bf C}^{n}$ and $w\in{\bf C}_{w}$ with the domain
${\bf C}^{n+1}={\bf C}^{n}\otimes{\bf C}_{w}$.
In principle, $w={\cal F}$ is expanded in the form of Hartogs series with a centre $a\in{\bf C}_{w}$:
\begin{eqnarray}
H(\{\mu_{f_{j}}\},\{\alpha_{j}\},z,\bar{z},\hbar,w) 
&=& \sum^{\infty}_{l=0} \alpha_{l}(\{\mu_{f_{j}}\},\{\alpha_{j}\},z,\bar{z},\hbar)(w-a)^{l}.
\end{eqnarray}
In this example, $0\le |\mu_{f_{j}}| <1$ ( $j=1,\cdots,n$ ) have to be satisfied
for making $f_{j}$ as quasiconformal, 
and the domain of the space of $(\mu_{f_{1}},\cdots,\mu_{f_{n}})$ becomes 
a poly-unit-disc $\bigotimes^{n}_{j=1}{\cal D}_{\mu_{f_{j}}}$ for the star product ${\cal F}$.
Moreover, for making ${\cal F}$ as quasiconformal, 
$0\le |\mu_{\cal F}|=|\sum^{n}_{j=1}\mu_{f_{j}}\alpha_{j}/\sum^{n}_{j=1}\alpha_{j}|<1$
has to be satisfied.
When, $f_{1}=f_{2}=\cdots=f_{n}$, the quantization cannot be performed.

\vspace{10mm}

For example, we consider the field $\phi\equiv Ne^{i\alpha z}e^{i\beta\bar{z}}$
which obey the bosonic statistics. $\phi$ is quasiconformal.
Then, one finds the Lagrange function densities as follows:
\begin{eqnarray}
{\cal L} 
&\equiv& \partial_{x}\phi^{\dagger}\partial_{x}\phi + \partial_{y}\phi^{\dagger}\partial_{y}\phi   
\nonumber \\
&=& 2\bigl( |\partial_{z}\phi|^{2} + |\partial_{\bar{z}}\phi|^{2} \bigr)   \nonumber \\
&=& |N|^{2}\Bigl\{ |\alpha+\beta|^{2}+|\alpha-\beta|^{2} \Bigr\}
e^{i(\alpha-\bar{\beta})z}e^{i(\beta-\bar{\alpha})\bar{z}}    \nonumber \\
&=& |N|^{2}\Bigl\{ |(1+\mu_{\phi})\alpha|^{2}+|(1-\mu_{\phi})\alpha|^{2} \Bigr\}
e^{i(\alpha-\overline{\mu_{\phi}}\bar{\alpha})z}e^{i(\mu_{\phi}\alpha-\bar{\alpha})\bar{z}},    \nonumber \\
{\cal L}_{\star} &\equiv& 
\partial_{x}\phi^{\dagger}\star \partial_{x}\phi + \partial_{y}\phi^{\dagger}\star \partial_{y}\phi   
\nonumber \\
&=& e^{-i\hbar(|\alpha|^{2}-|\beta|^{2})}{\cal L}
= e^{-i\hbar(1-|\mu_{\phi}|^{2})|\alpha|^{2}}{\cal L}  \nonumber \\
& & ( \partial_{\bar{z}}\phi = \mu_{\phi}\partial_{z}\phi ).
\end{eqnarray}
Hence in this model, the deformation quantization only gives ${\cal L}$ the phase factor 
$e^{-i\hbar(1-|\mu_{\phi}|^{2})|\alpha|^{2}}$, and it is controlled by the Beltrami
coefficient $\mu_{\phi}$.

\vspace{10mm}

By utilizing the Fuchsian model, Bers embeddings, Weil-Petersson metrics, and the methods
of complex analysis of several variables, the construction of a Teichm\"{u}ller theory
for the star product would be done. 
A generalization of our results to the case of noncommutative ring ${\bf C}$ with $[z,\bar{z}]\ne 0$ 
is also an interesting problem~[8].
These problems will be examined in forthcoming papers by the author.

\vspace{10mm}

Finally, we give a short comment.
It was shown that, the star product has a deep connection with
topological field/string theory ( the Poisson $\sigma$-model )~[6,12].
By the examinations of quasiconformal mappings of the star products
and Teichm\"{u}ller spaces, the deformation quantization of the star products
would obtain relations with complex dynamics and fractals.
Hence, our work might make a road for introducing 
a concept of complex dynamics into topological field/string theories.

\end{document}